\documentclass[apj]{emulateapj} 
\usepackage{natbib}
\usepackage{graphicx}
\usepackage{color}
\usepackage{url}
\usepackage[backref, breaklinks, colorlinks,citecolor=blue]{hyperref}

\shorttitle{ M/L-[Fe/H] relation in M31}

\shortauthors{A. H. Zonoozi,  H. Haghi and P. Kroupa }
\begin{document}

\title{{\bf  A Possible Solution for the M/L-[Fe/H] Relation of Globular Clusters in M31: A metallicity and density dependent top-heavy IMF} }

\author{A. H. Zonoozi \altaffilmark{1},  H. Haghi \altaffilmark{1}, P. Kroupa \altaffilmark{2}}
\email{a.hasani@iasbs.ac.ir}  \email{haghi@iasbs.ac.ir}  \email{pavel@astro.uni-bonn.de}
\altaffiltext{1}{Institute for Advanced Studies in Basic Sciences (IASBS), P. O. Box 45195-1159, Zanjan, Iran}
\altaffiltext{2}{HISKP, Bonn, Germany}

\begin{abstract}

The observed mass-to-light ($M/L$) ratios of a large sample of GCs in M31 show an inverse trend with metallicity compared to what is expected from Simple Stellar Population (SSP) models with an invariant canonical stellar IMF, in the sense that the observed $M/L$ ratios decrease with increasing metallicity. We show that incorporating the effect of dynamical evolution the SSP models with a canonical IMF can not explain the decreasing  $M/L$ ratios with increasing metallicity for the M31 GCs. The recently derived top-heavy IMF as a function of metallicity and embedded cluster density is proposed to explain the lower than expected $M/L$ ratios of metal-rich GCs. We find that the SSP models with a top-heavy IMF, retaining a metallicity- and cluster mass- dependent fraction of the remnants within the clusters, and taking standard dynamical evolution into account can successfully explain the observed $M/L-[Fe/H]$ relation of M31 GCs. Thus we propose that the kinematical data of GCs can be used to constrain the top-heaviness of the IMF in GCs.

\end{abstract}

\section{Introduction}\label{Sec:Intro}

Globular cluster (GC) systems are major tracers for studying the formation and early evolution of their  host galaxies. They are also useful for studying the stellar initial mass function (IMF) because they provide us with large nearly single-aged and mono-metallicity populations of stars located at the same distance which means that the only significant difference between the stellar populations in different clusters is the mass and metallicity of the clusters.

The stellar mass-to-light ($M/L$) ratio is an important parameter of a GC as it establishes a connection between the luminous and gravitating mass of a population. The bolometric luminosity of a GC is mostly due to the small number of evolved stars, while the mass is dominated by the more numerous unobserved low-mass stars and possibly stellar remnants. This implies that the M/L ratio of a GC reflects its present-day mass function (MF), and hence through its evolution also the IMF.

For a simple stellar population, in which all stars are formed in a single-metallicity instantaneous burst, the stellar $M/L$ ratio depends on the age and the IMF. In the beginning, a large fraction of the mass in the cluster can be found in high-mass stars (i.e., $M\geq M_{\odot}$).
As a result of stellar evolution, the fraction drops and the high-mass end of the mass function turns into compact remnants. The total luminosity of a star cluster which is due to the high-mass stars drops by an order of magnitude within the  first 2 Gyr and by roughly another order of magnitude within the next 10 Gyr of evolution \citep{Baumgardt03}. Therefore, the $M/L$ ratio increases constantly as the most luminous stars disappear.

While stellar evolution only raises the $M/L$ ratio of a stellar population with time, on the other hand, dynamical evolution leads to a decrease of the $M/L$ ratio because of the preferential two-body relaxation driven evaporation of faint, low-mass stars (i.e., high $M/L$ stars) for star clusters evolving in a tidal field of a host galaxy.

Within the last decade the structural parameters and kinematical properties of extragalactic GCs with distances up to a few tens of Mpc have been well measured. For example, Strader et al. (2009, 2011) presented the observed velocity dispersions for 200 GCs in M31 using new high-resolution MMT/Hectochelle spectra and covering a wide range of cluster masses and metallicities. For 163 GCs of this sample they have derived the $M/L$ values in both the optical (V-band) and near infrared (K-band).  Since line blanketing increases with metallicity, stellar population synthesis predicts fainter optical luminosities for more metal-rich clusters, while there is no similar dependence expected in the K-band. They found that the M/L ratios are lower than what is expected from the stellar population model predictions with a canonical stellar IMF. The discrepancy between the observed $M/L_V$ and the SSP model  is larger at high metallicities where the $M/L_V$ values fall below the SSP values, by a factor of more than three. Moreover, this discrepancy is more pronounced for clusters with a lower mass and for clusters with a shorter two-body relaxation time scale (Strader et al. 2009).


There are two ways to reduce $M/L$: adding stars with low $M/L$ ratios (i.e., $RGB/AGB$ stars), or removing stars with high  $M/L$ ratios (low-mass dwarfs).  Different IMFs for metal-poor and metal-rich clusters are proposed by Strader et al. (2009) to explain the $M/L-[Fe/H]$ relation of M31 GCs.

Most recently, Shanahan \& Gieles (2015) showed that the mass segregation bias in $(M/L)_V$ as a function of [Fe/H] can explain the observed discrepancy in $M/L$ between the dynamical and SSP models, without invoking the depletion of low-mass stars, or variations in the IMF. This scenario is supported by the discrepancy between the observed $(M/L)_V$ and the SSP models being more pronounced for low-mass clusters with a shorter two-body relaxation timescale (Strader et al. 2011), because  dynamical mass segregation is a more efficient process in clusters with lower mass.  However, ignoring the effect of two-body relaxation driven evaporation of low-mass stars is nearly valid only in clusters with a long two-body relaxation time-scale, and for GCs that are older than their half-mass relaxation time-scale the depletion of low-mass stars should be included.


Despite all the evidence for a universal IMF (e.g., Kroupa 2001; Kroupa 2002; Bastian et al. 2010), recently several indications have begun emerging for a possible environment dependency of the shape
of the IMF. Several observational and theoretical indications suggest the IMF to become top-heavy under extreme
starburst conditions (Marks et al. 2012, Dabringhausen et al. 2009, 2010, 2012; see Kroupa et al. 2013 for a review on the evidences for the top-heavy IMF). The data suggest the IMF to become less top-heavy with increasing cluster metallicity and decreasing density.
The aim of the present paper is to assess the different scenarios for explaining the $M/L-[Fe/H]$ discrepancy and to constrain the most reliable of them. The paper is organized as follow: In Sec. 2 we investigate if only dynamical evolution is able to explain this discrepancy. In Sec. 3 we assess whether SSP models using the top-heavy IMF can explain the $M/L-[Fe/H]$ correlation. Our conclusion and discussion are presented in Sec. 4

\section{The canonical IMF: dynamical evolution and the mass-to-light ratio}\label{DynEvol}

\begin{figure*}
\centering
\includegraphics[width=80mm]{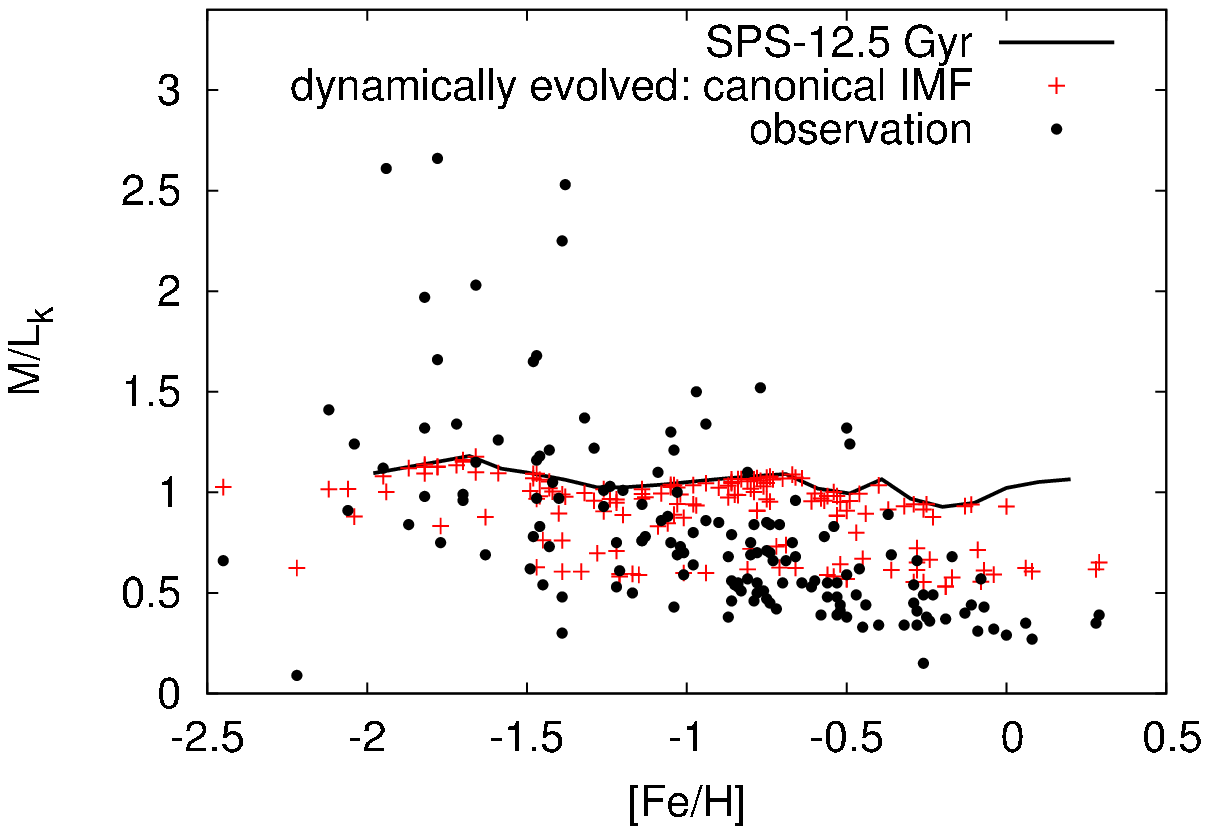}
\includegraphics[width=80mm]{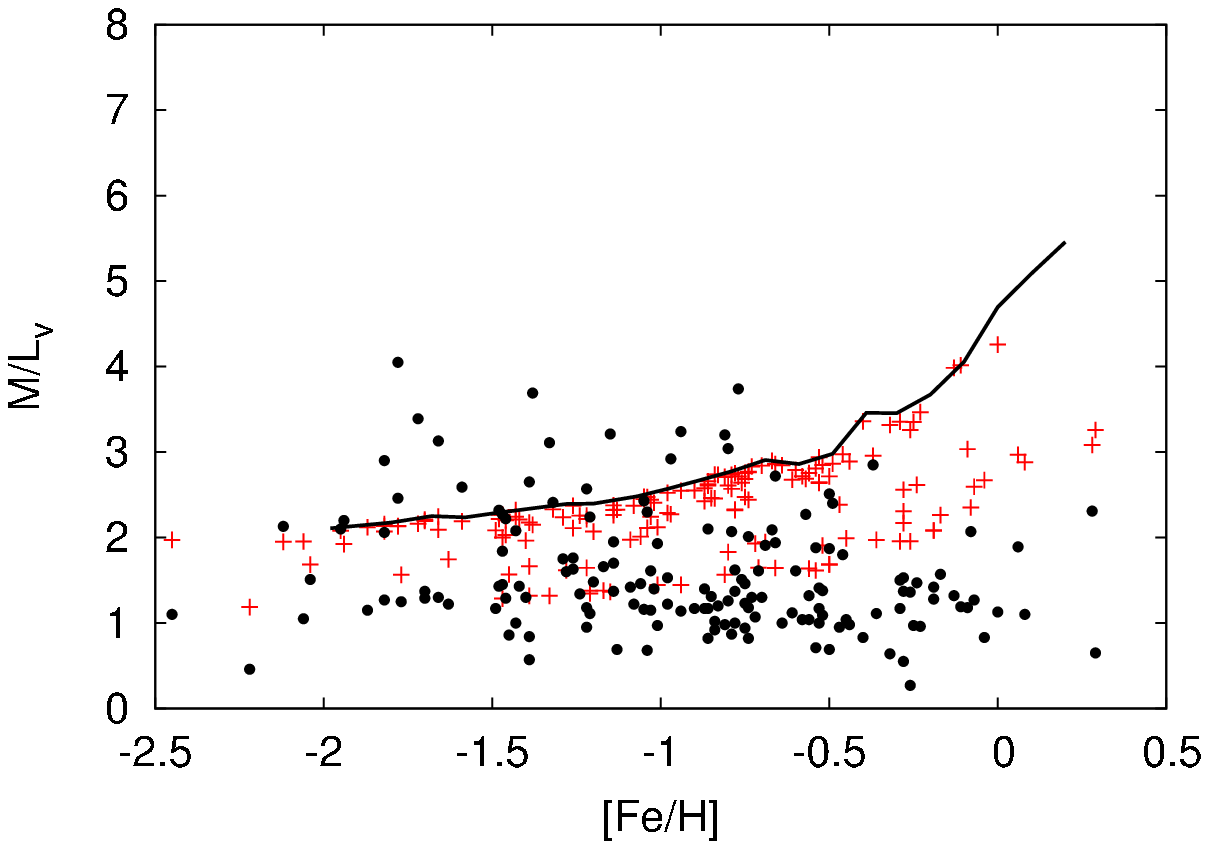}
\caption{Mass-to-light ratio vs. [Fe/H] in the K- (left panel)and V-band (right panel) for M31 GCs taken from Strader et al (2011, black dots).  A SSP model curve for a canonical IMF and an age of T=12.5 Gyr is overploted as a solid line in each panel. The $M/L$ values of the metal rich GCs are significantly lower than expectation from SSP models.  Both optical and near-infrared $M/L$ ratios of M31 GCs decrease with increasing metallicity, in contrast to the SSP predictions.  This trend is even more evident in the K-band, where no strong dependence of $M/L_K$ on metallicity is expected in the SSP models based on an invariant canonical IMF. The  M/L ratios, calculated at the $[Fe/H]$ values of the black dots with initial ratios as given by the SSP models and corrected for dynamical evolution over 12.5 Gyr are shown as red symbols.  Although the preferential evaporation of low-mass faint stars as a result of mass segregation and overflow over the tidal boundary of low-mass stars leads to a decrease of the $M/L$ ratios,  the observed  trend is still unexplainable  only by  dynamical evolution. For details see text.}
\label{dynamic}
\end{figure*}

Figure \ref{dynamic} shows the observed $M/L$ ratio of M31 GCs versus their metallicity [Fe/H] in the V- and K-bands (Strader et al. 2011). In order to compare with the theoretical expectation from SSP models, we overplot the SSP models for both optical and near-infrared $M/L$ ratios assuming an age of 12.5 Gyr and the canonical  IMF using the flexible stellar population synthesis  code (FSPS, Marigo et al. 2008, Conroy et al 2009; Conroy and  Gunn 2010),  in which the Padova isochrones and the Basell stellar library are utilized.  The canonical ($\alpha_2=\alpha_3$) IMF, $\xi(m)$, can be conveniently  written as a two-part power-law  function, $\xi(m) \propto m^{-\alpha_i}$, where $\alpha_1=1.3$ for stars with mass $0.08\leq m/M_{\odot}\leq0.5$, $\alpha_2=2.3$  for $0.5\leq m/M_{\odot}\leq1$ and  $\alpha_3$ for $m>1 M_{\odot}$ (Kroupa 2001, Kroupa et al. 2013) with the upper mass-limit of 100 $M_{\odot}$. As can be seen in Figure \ref{dynamic} (solid line), no strong dependence of $M/L_K$ on metallicity is expected for SSP models, while $M/L_V$ increases significantly with metallicity.  An observed trend of strongly decreasing $M/L_K$ values with increasing metallicity for M31 GCs can be seen, while only a minimal dependency of $M/L_K$ values on metallicity is expected from SSP models (e.g., Bruzual \& Charlot 2003; Maraston 2005; Conroy \& Gunn 2010). The observed $M/L_V$ ratios of metal-rich GCs deviate strongly from the  SSP models which predict fainter luminosities and  larger $M/L_V$ ratios for more metal-rich systems.

The stellar population models only account for the effects of stellar evolution, whereas  dynamical evolution and preferential depletion of low mass (i.e., high $M/L$) stars through two-body relaxation, mass segregation and evaporation from evolving GCs changes the shape of the stellar MF within a cluster \citep{Vesperini97, Baumgardt03}, causing the $M/L$ ratio to decrease.


In order to take the dynamical evolution into account we follow the N-body results of Baumgardt \& Makino (2003) and estimate the effect of dynamical evolution
on the stellar MF-slope of M31 GCs.  The dissolution time of star clusters on circular orbits can be expressed as (Baumgardt \& Makino 2003):

\begin{equation}
T_{diss} = \beta \left[ \frac{N}{\ln(\gamma N)} \right]^{x}\frac{R_G}{[kpc]}\left(\frac{V_G}{200 km/sec} \right)^{-1},
\end{equation}
where $\gamma=0.02$ (Giersz \& Heggie 1996),  $x=0.75$ and $\beta=1.91$  for King $W_0=5.0$ clusters (Baumgardt \& Makino 2003). The initial number of stars, $N$, is calculated from the initial mass of clusters assuming a mean stellar mass of $0.55 M_\odot$. It should be noted that the exact value of the mean mass depends on the assumed IMF. In addition, it is varying in the simulated clusters because the mass function changes during the evolution.  But, the change is only marginal and does not affect the nature of the conclusions.

In order to iteratively calculate the initial mass of individual clusters (using eqs. 10 and 12 of Baumgardt \& Makino 2003) to produce a cluster with the present-day mass after 12.5 Gyr, one needs the orbital parameters of each cluster which are not yet available. To estimate the the initial masses of the clusters we refer to table 2 of Baumgardt \& Makino (2003) where 10 clusters are listed for which orbital information and determinations of the slope of the mass function are simultaneously available. According to their table 2 the mean value of the initial mass over the current mass is $<M/M_0> \simeq 3 \pm 1.5$, where $M$ is the present-day mass (as given in Table 4 of Strader et al. 2011) and $M_0$ is the initial mass of the cluster. We assume a mean value independent of $M_0$ because, although more massive clusters evaporate their stars on a longer time scale than low-mass clusters, if it were the case that the initially more massive clusters were more centrally concentrated in the proto-galaxy therewith being subject to a stronger tidal field, this bias would be reduced. For example, Arp~220 has a central star-burst in very massive forming clusters with an abnormally high supernova rate there (Dabringhausen et al. 2012 and references therein); our own Milky Way has very massive and very young central clusters (Arches and Quintuplet, e.g. Portegies Zwart et al. 2010); the nearby disk galaxy M33 has been suggested to have a radially systematically decreasing most-massive young star-cluster population (Pflamm-Altenburg et al. 2013); and many galaxies have central nuclear clusters. Therefore we assume for this analysis that the above  mean value is independent  of $M_0$. For each individual observed cluster we assume that $M/M_0$ is a random number which is generated from a Gaussian distribution with a mean value of 3 and standard deviation of 1.5.

Since only the projected galactocentric radius of each M31 GC is available (Caldwell et al 2009), calculating the dissolution time and consequently the dynamical evolution of a cluster in a realistic tidal field is uncertain. Assuming clusters move on eccentric orbits with an eccentricity of $e$, the life times of clusters, $T_{diss} (e)$,  decrease as a function of eccentricity compared to  clusters moving on circular orbits with radius equal to the apogalactic distance  of the eccentric orbit as

\begin{equation}
T_{diss} (e) = T_{diss} (0) (1 -e).
\end{equation}

Since the orbital shapes of M31 GCs are unknown, we assume that these clusters are following eccentric orbits with an average eccentricity of $e=0.5$. We also assume they are likely near the apocenter to estimate the maximum effect of dynamical evolution.
It has been shown that the slope of the stellar mass function at time $T$ and in the low mass range, $m\leq 0.5 M_{\odot}$, changes in a universal way, independent on the orbital shape of the cluster as (Baumgardt \& Makino 2003)
\begin{eqnarray}
\nonumber \alpha_{1}&=& 1.3- 1.51 \left(\frac{T}{T_{diss}(e)}\right)^{2}+ 1.69\left(\frac{T}{T_{diss}(e)}\right)^{3}\\
 \label{eq2}&&~~~~~~~~-1.50 \left(\frac{T}{T_{diss}(e)}\right)^{4}.
\end{eqnarray}

This is used here to estimate the impact of dynamical evolution on the stellar $M/L$ ratio of each individual GC assuming $T=12.5$ Gyr. Note that the change in the slope of the mass function in the mass range $m\geq 0.5 M_{\odot}$  ($\alpha_2$) over time is negligible and is more noticeable at the low-mass part ($\alpha_1$).  This is because of the dynamical mass segregation that the high-mass stars sink into the inner region of the cluster and experience a weak tidal field, while the low-mass stars that are distributed in the outer regions preferentially escape due to the external tidal interaction (see e.g., Baumgardt \& Makino 2003 and Haghi et al. 2015 ).

Figure \ref{dynamic} depicts the observed $M/L$ ratios of M31 GCs and the prediction of SSP models with (red plus symbols) and without (solid lines) adding the effect of dynamical evolution. To calculate the position of the red symbols we first assume the initial (i.e. birth) $M/L$ ratio of an observed cluster (black dots in Fig. \ref{dynamic}) at its $[Fe/H]$ value is that provided by our SSP model shown by the solid curves in Fig. \ref{dynamic}. These $M/L$ values correspond to the canonical IMF (Sec.2) and are evolved to the $M/L$ values for the present-day MF using Eq. 3. The $M/L$ value at $T=12.5\,$Gyr is then calculated anew using the SSP code FSPS for the so-obtained present-day MF. Thus, if the model captures reality well, then the so-calculated $M/L$ values (the red crosses) ought to lie close to their corresponding black dots.

Basically we expect dynamical evolution to decrease the $M/L$ ratios, but this decline is not metallicity dependent and does not show any trend of the $M/L$ ratios with $[Fe/H]$, neither in the $V-$ nor in the $K$-band. In other words, although the dynamical evolution leads to a decrease of the stellar $M/L$ ratios from what is predicted by SSP models, it does not provide the observed trend between $M/L$ and metallicity. We therefore conclude  that standard dynamical evolution alone is not the reason for the  observed discrepancy between the observed data and the SSP models, confirming the conclusion by Strader et al (2011).

\section{The top-heavy IMF case} \label{ssec:Top}

\begin{figure*}
\centering
\includegraphics[width=80mm]{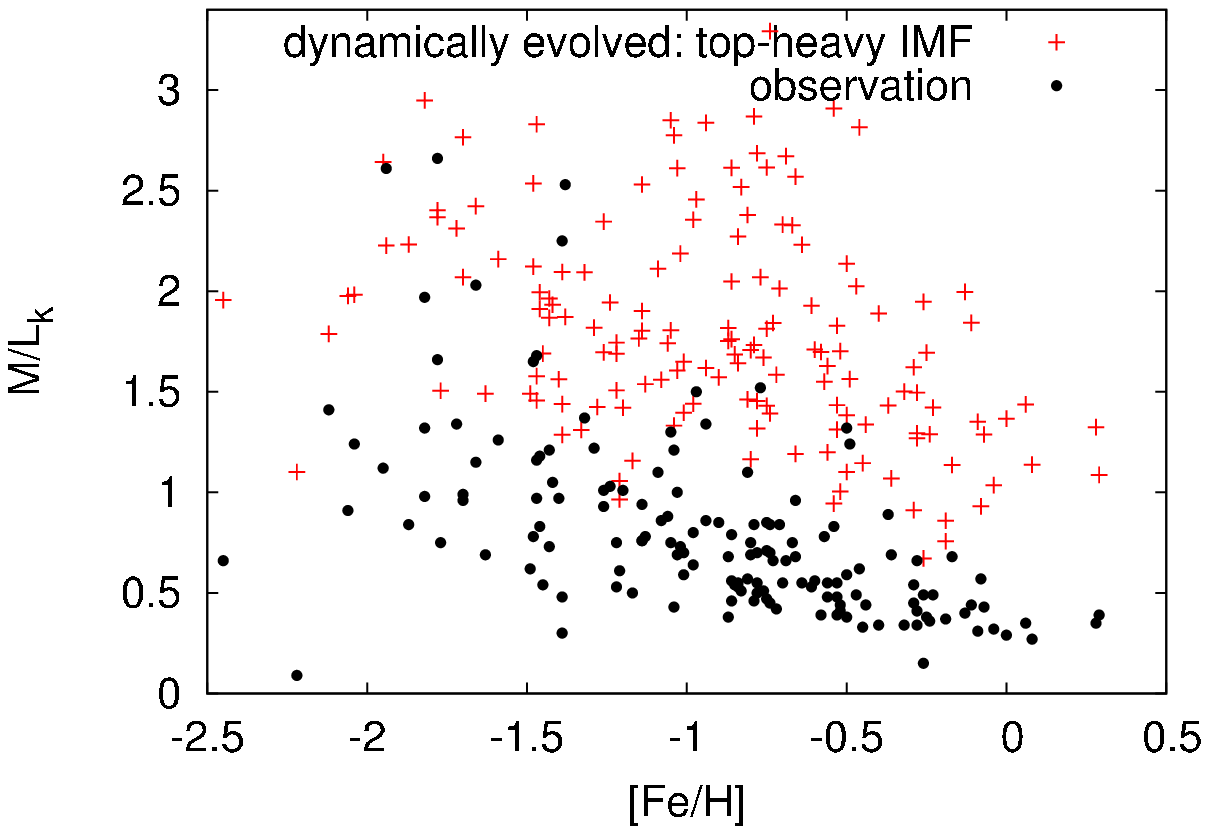}
\includegraphics[width=80mm]{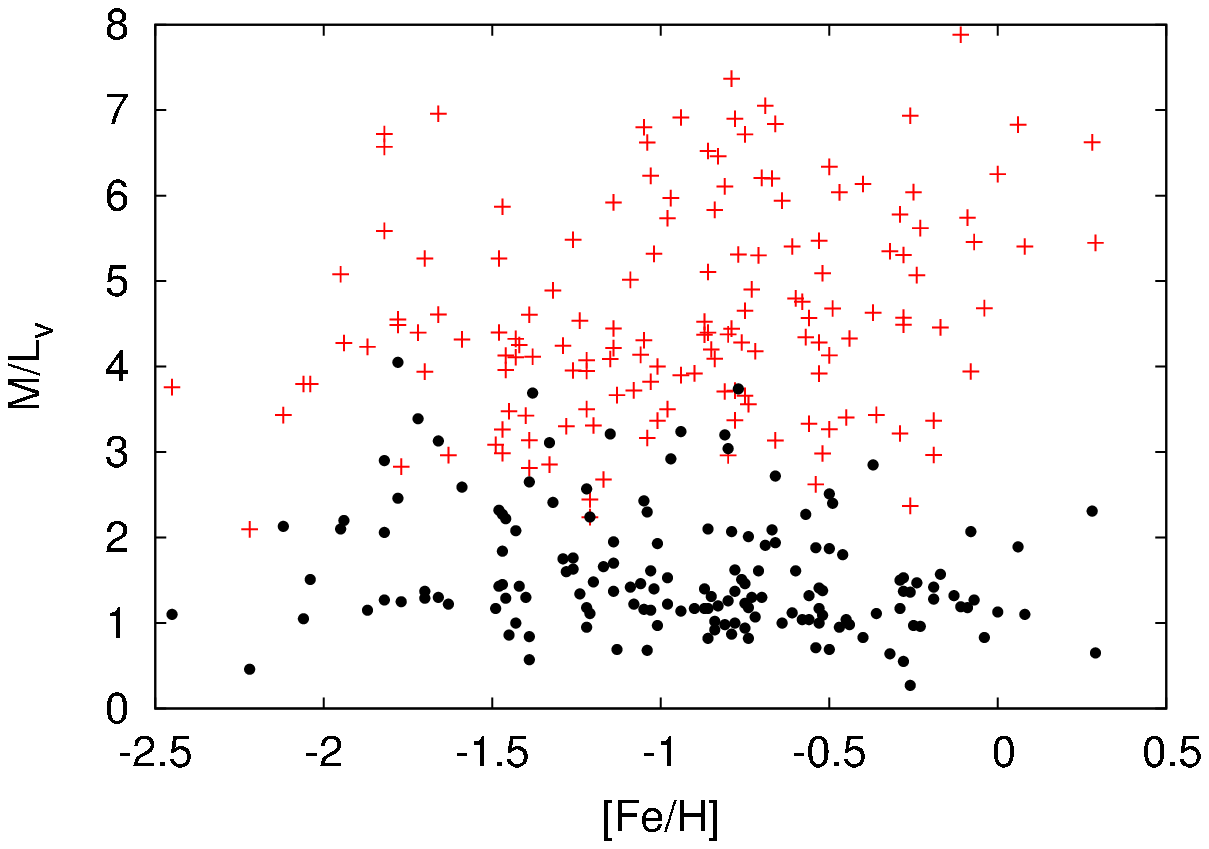}
\caption{$M/L$ vs. $[Fe/H]$ in the K- and V-band for M31 GCs (black dots). The SSP prediction for each cluster assuming a top-heavy IMF (Eq. 4) and corrected for dynamical evolution but retaining  all stellar remnants is shown as the red plus symbols (see Sec. 2 and 3 for details).  For the model to reproduce the data, the model present-day values (the red crosses) should be distributed as the black dots. This is better here than in Fig.1, but still not sufficient.} The present-day masses of individual clusters are taken from Strader et al (2011).
\label{top}
\end{figure*}


\begin{figure*}
\centering
\includegraphics[width=80mm]{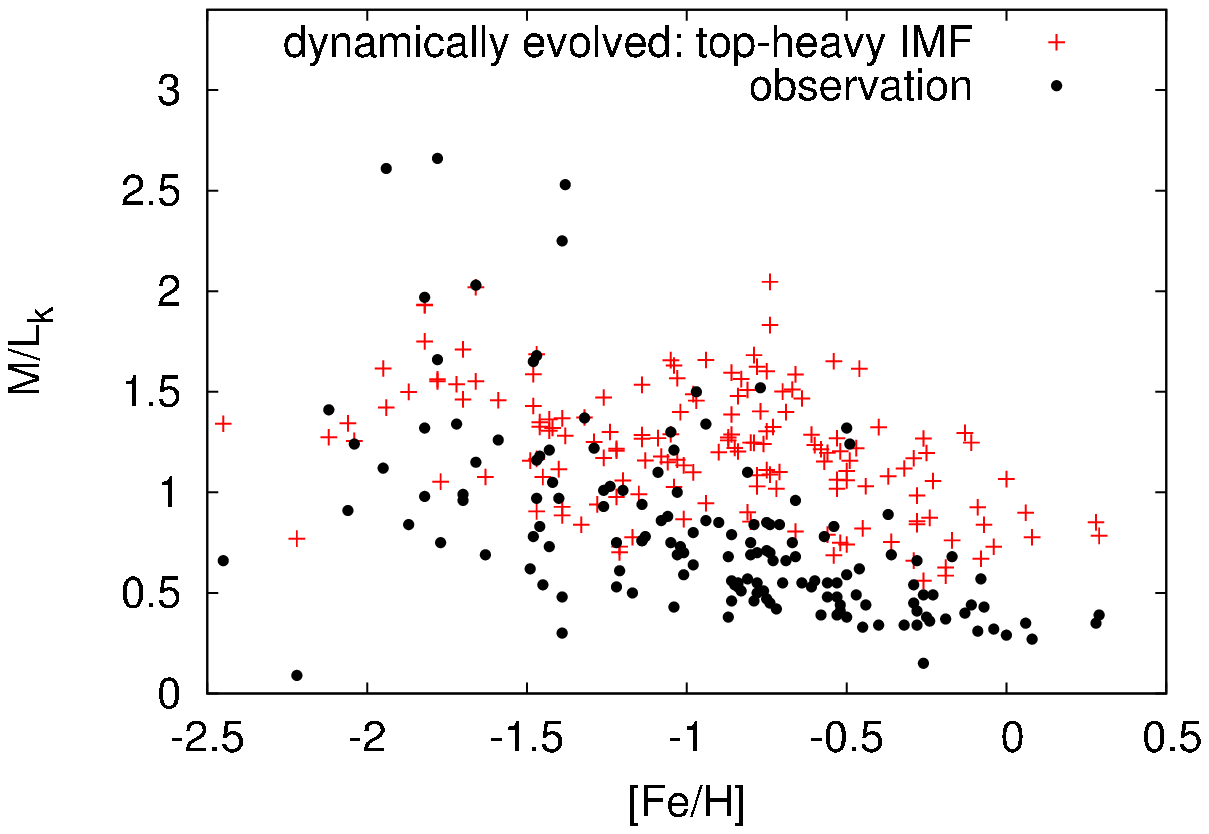}
\includegraphics[width=80mm]{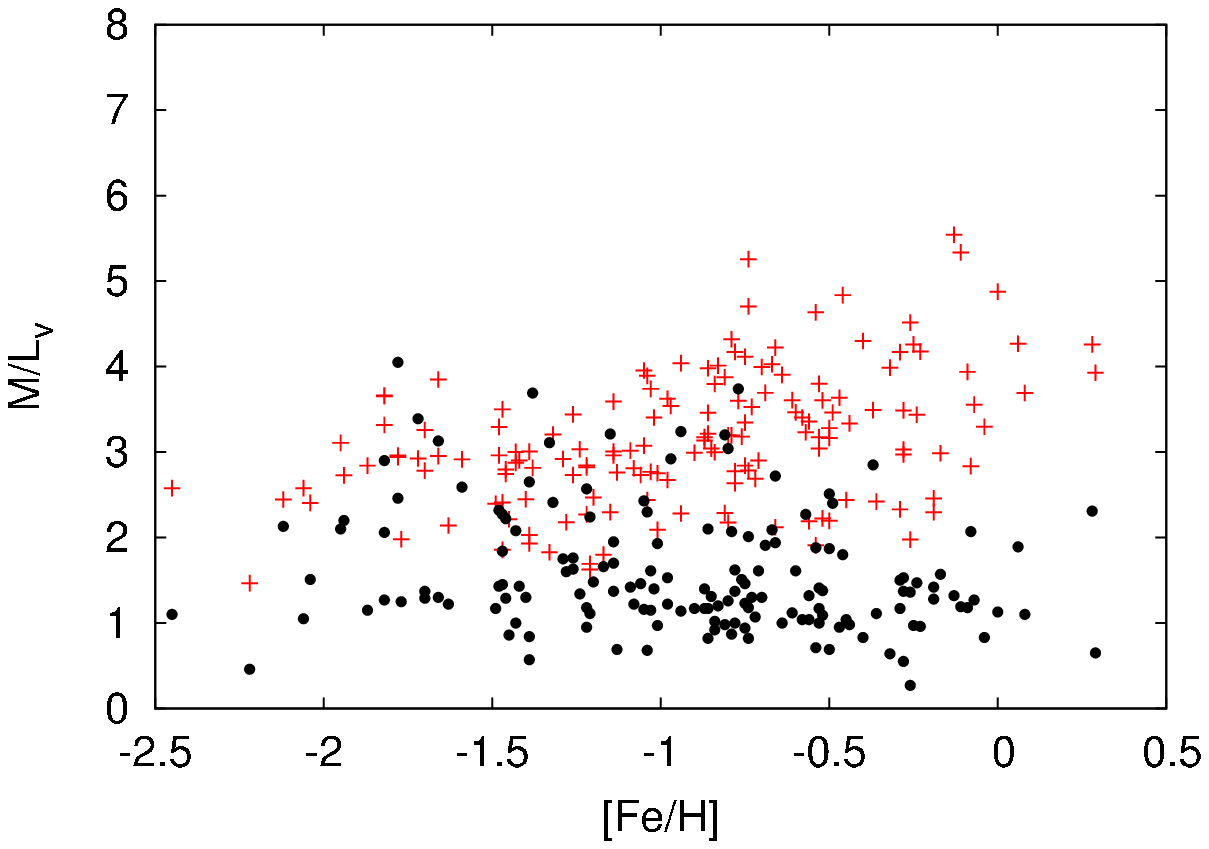}
\caption{The same as Fig 2, but keeping 30\%  of all black holes (BHs) and neutron stars (NSs). }
\label{top-30}
\end{figure*}

In a gas cloud of low-metallicity the Jeans mass is larger favoring the formation of more massive stars and the fraction of high-to-low mass stars increases (Larson 1998). Adams \& Fatuzzo (1996) on the other hand discuss the possibility that because there is no single Jeans mass per cloud, a stellar mass grows to a final mass through a balance between accretion rate and self-generated energy feadback, both of which are metallicity and density dependent with the same general metallicity dependence of the stellar mass as in the above Jeans-mass argument.  In very dense star-forming cores, pre-stellar cores may coalesce before they form proto-stars thus leading to a top-heavy IMF (Dib et al. 2007, Dib, Kim, \& Shadmehri 2007).

These fundamental theoretical arguments  lead to the expectation that the IMF ought to become  metallicity dependent by being  more top-heavy (i.e., flatter) in metal-poor and denser environments. It was disconcerning  that until recently observational data did not indicate this long-expected IMF variation   (Kroupa et al. 2013).

Indeed, the inferred IMF slope at the high-mass end, $\alpha_3$, for a sample of MW GCs  suggests that the high-mass IMF was more top-heavy (flatter) in more massive and denser environments (see figs 2 and 3 of Marks et al. 2012). This implies that denser and to a lower extend metal-poorer systems form more massive stars compared to the canonical IMF (see also Weidner et al 2013 and Kroupa et al. 2013 for  reviews on the evidence for the top-heavy IMF). The independent data analyzed by Dabringhausen et al. (2009, 2012) and Marks et al. (2012) on ultra-compact dwarf-galaxies and globular clusters, respectively,  suggest that the slope of the IMF for stellar masses larger than $1 M_{\odot}$, $\alpha_3$,  and its variation can been described as follows

\begin{equation}\label{alpha3}
     \alpha_{3}= \left\{
        \begin{array}{lcl}
        +2.3~~~~~~~~~~~~~~~~~~~~ , x<-0.87, \\
        -0.41\times x+1.94~~~~, x\geq-0.87,
        \end{array}
        \right\}
\end{equation} 

\noindent where $x=-0.14 [Fe/H]+0.99\log_{10}(\rho_{ cl}/(10^{6} M_\odot pc^{-3}))$.
Here in this section we reproduce the SSP analysis with assuming a density and a metallicity dependent IMF such that the IMF is more top-heavy for lower metallicity and larger density (Eq. 4).  In order to calculate the pre-GC cloud-core density, $\rho_{cl}=3M_{cl}/4\pi r_h^3$, one needs to estimate the original molecular cloud core mass in gas and stars, $M_{cl}$, and the half-mass radius. Assuming a star formation efficiency of 33\% \citep{Lada03, Megeath16} the mass of the original molecular cloud core is three times the mass of the embedded cluster $M_{ecl}$ (i.e., $M_{cl}=3M_{ecl}$). This star formation efficiency is remarkably consistent with very detailed $N$-body modelling of well observed very young star clusters ranging in mass from $10^3$ to $10^5 M_{\odot}$ \citep{kroupa01b, Banerjee13, Banerjee14}. The birth half-mass radius, $r_h$, can be calculated as follow from an analysis of star-forming systems by Marks \& Kroupa (2012)

\begin{equation}
r_h (pc) = 0.1\times\left(\frac{M_{ecl}}{M_{\odot}}\right)^{0.13}.
\end{equation}


This birth half-mass radius is used only to calculate $\rho_{cl}$, and we assume that the young GCs thereafter evolves through stellar-evolution mass loss and expulsion of residual gas to the present-day radii. The K- and V-band $M/L$ ratios from the SSP models, calculated as in Sec. 2 but here assuming a top-heavy IMF and corrected for the effect of dynamical evolution but retaining all stellar remnants in the $T=12.5$ Gyr cluster model are compared to the observed values in Figure \ref{top}. Although, the top-heavy IMF now leads to a $M/L-[Fe/H]$ trend (especially in the K-band, compare with Fig. \ref{dynamic}), the predicted $M/L$ values (red crosses) are still larger than the observed values (black dots). As can be seen, even the effect of dynamical evolution can not sufficiently reduce the $M/L$  ratios to  become consistent with observations, and there remain a large number of GCs with observed $M/L$  ratios that are lower than the predicted ones by SSP models with a top-heavy IMF.

It should be noted that so far we kept 100\% of the stellar remnants within the clusters which leads to higher $M/L$  ratios, because the remnants contribute to the mass without any contribution to the luminosity. However, it is still unclear  how many remnants receive a velocity kick at formation and get ejected immediately. Some  recent studies have shown that even if black hole (BH)-formation kicks are not sufficiently high to eject BHs from young GCs a significant fraction of the formed BHs are expelled through stellar-dynamical evolution up to the typical ages (12 Gyr) of the GCs \citep{Mackey08, Banerjee10, Banerjee11, Breen13, Heggie14,  Sippel13, Morscher15}. This implies that the assumption of a 100\% retention fraction is not reasonable for GCs.

If BHs receive similar velocity kicks as neutron stars (NSs) upon formation, and if not all NSs  receive a high kick, a retention fraction of 30\% for all remnants may be assumed. Keeping 30\% of all BHs and NSs within the clusters we recalculate the $M/L$  ratios from SSP models with a top-heavy IMF. As shown in Fig. \ref{top-30}, the improvement with respect to the panels in Fig. \ref{dynamic} is considerable. The use of the top-heavy IMF in SSP models, and of the dynamical evolution and partially keeping the remnants in GCs contribute to this improvement.

\begin{figure*}
\centering
\includegraphics[width=80mm]{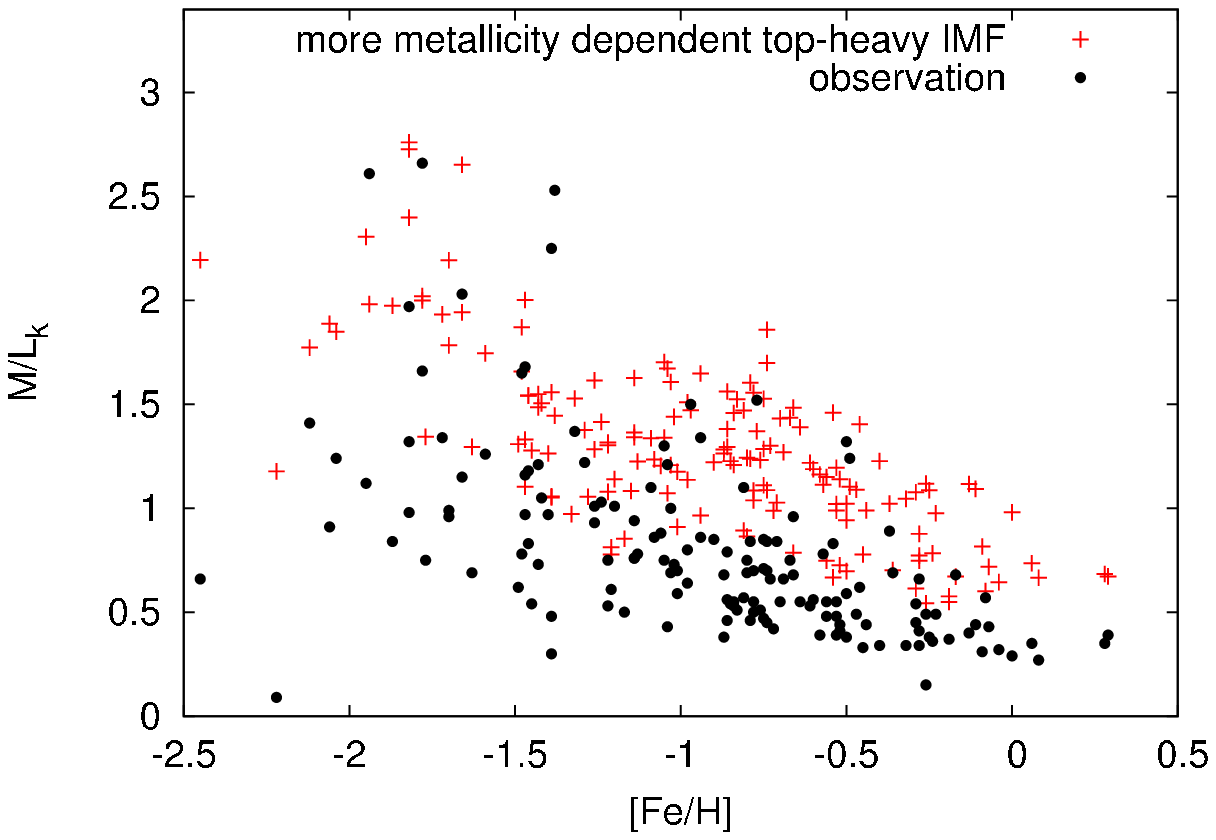}
\includegraphics[width=80mm]{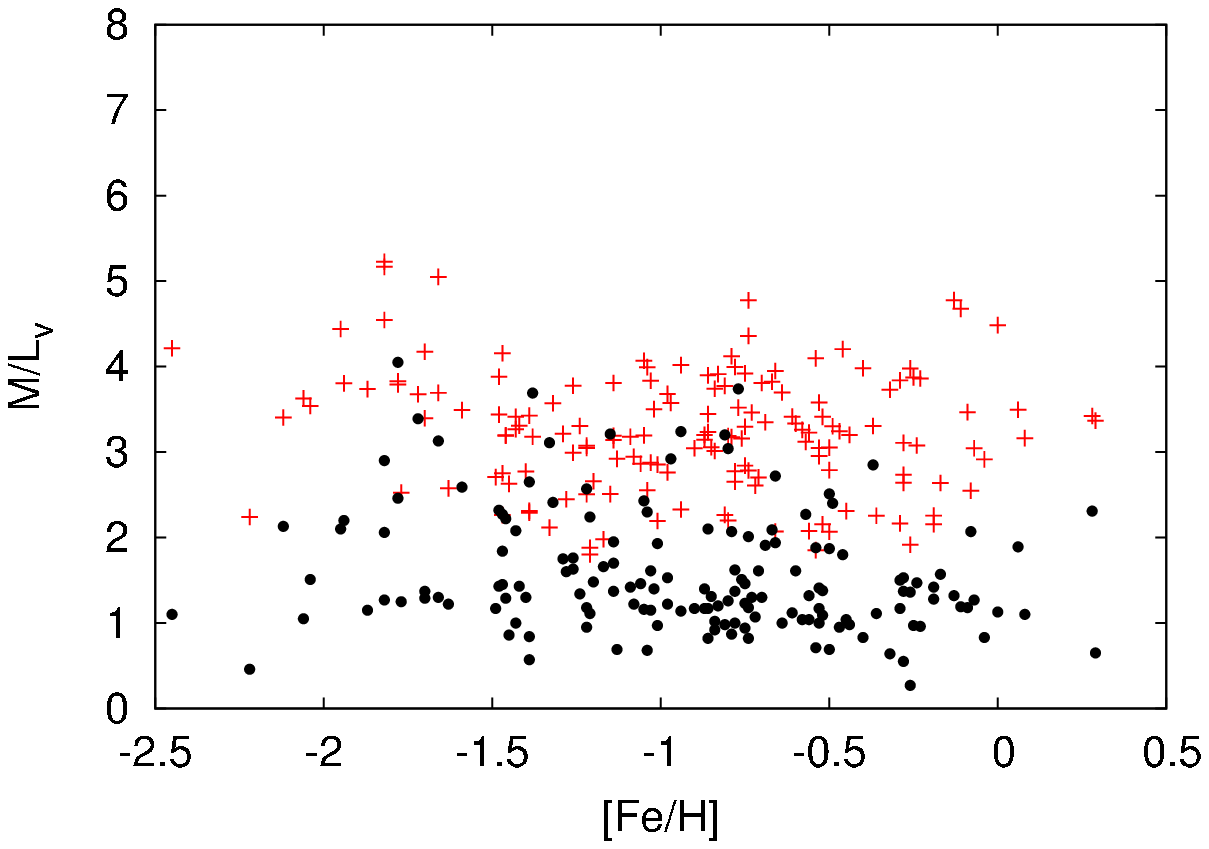}
\caption{The same as Fig.  \ref{top}, but assuming the stronger dependency of $\alpha_3$ on metalicity (Eq. 6), and 10\% retention fraction for stellar remnants. See text for more details. } \label{top-strong}
\end{figure*}

If the dependency of the IMF top-heaviness on the metallicity of the progenitor molecular cloud is enhanced, e.g., by changing the $x$ parameter  in Eq. 4 to

\begin{equation}
x=-0.70 [Fe/H]+0.99\log_{10}\left(\frac{\rho_{ cl}}{10^{6} M_\odot pc^{-3}}\right),
\end{equation}

\noindent a remarkable agreement between the observed  $M/L-[Fe/H]$ trends and that expected from theoretical models can be achieved as we show in Fig. \ref{top-strong}. Such an enhanced metallicity dependent top-heavy IMF may be argued for, given that the original formulation (Eq. 4) can at present only be seen as a first approximation as derived from GCs by Marks et al (2012).

\begin{figure*}
\centering
\includegraphics[width=80mm]{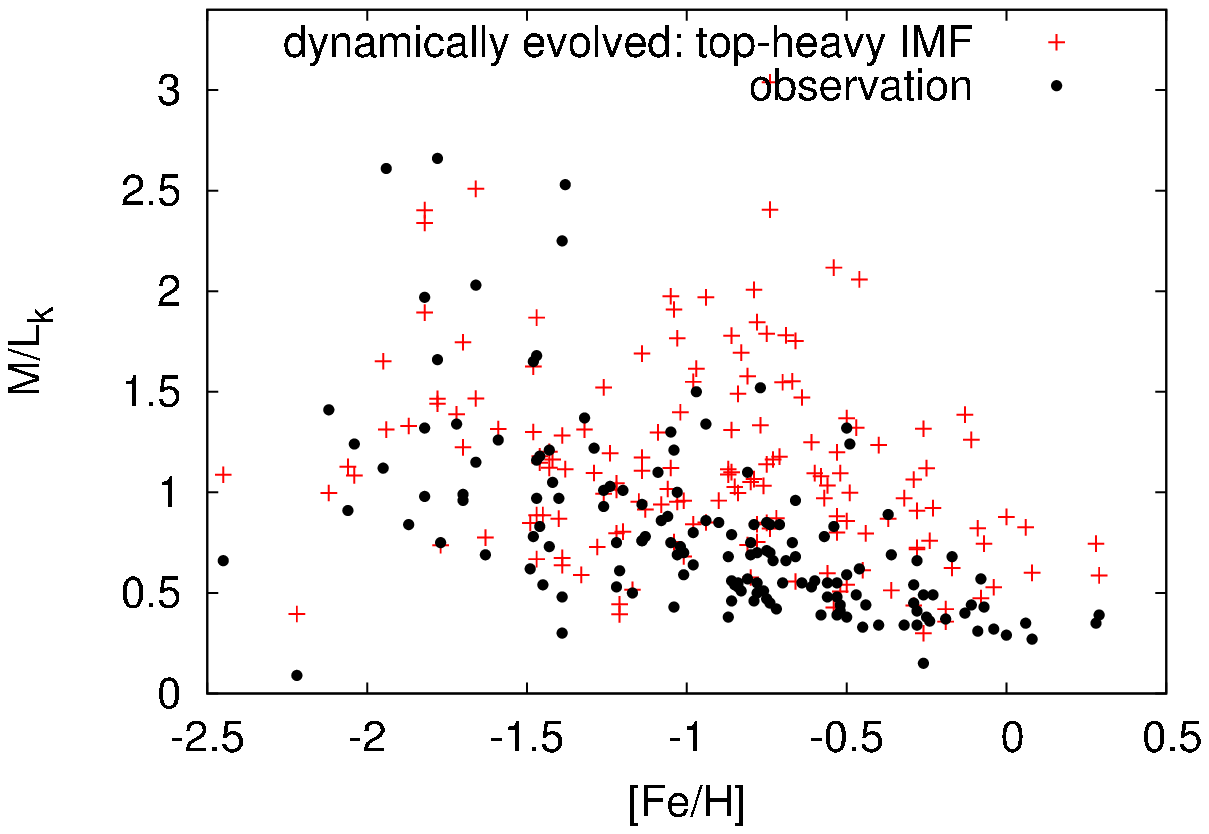}
\includegraphics[width=80mm]{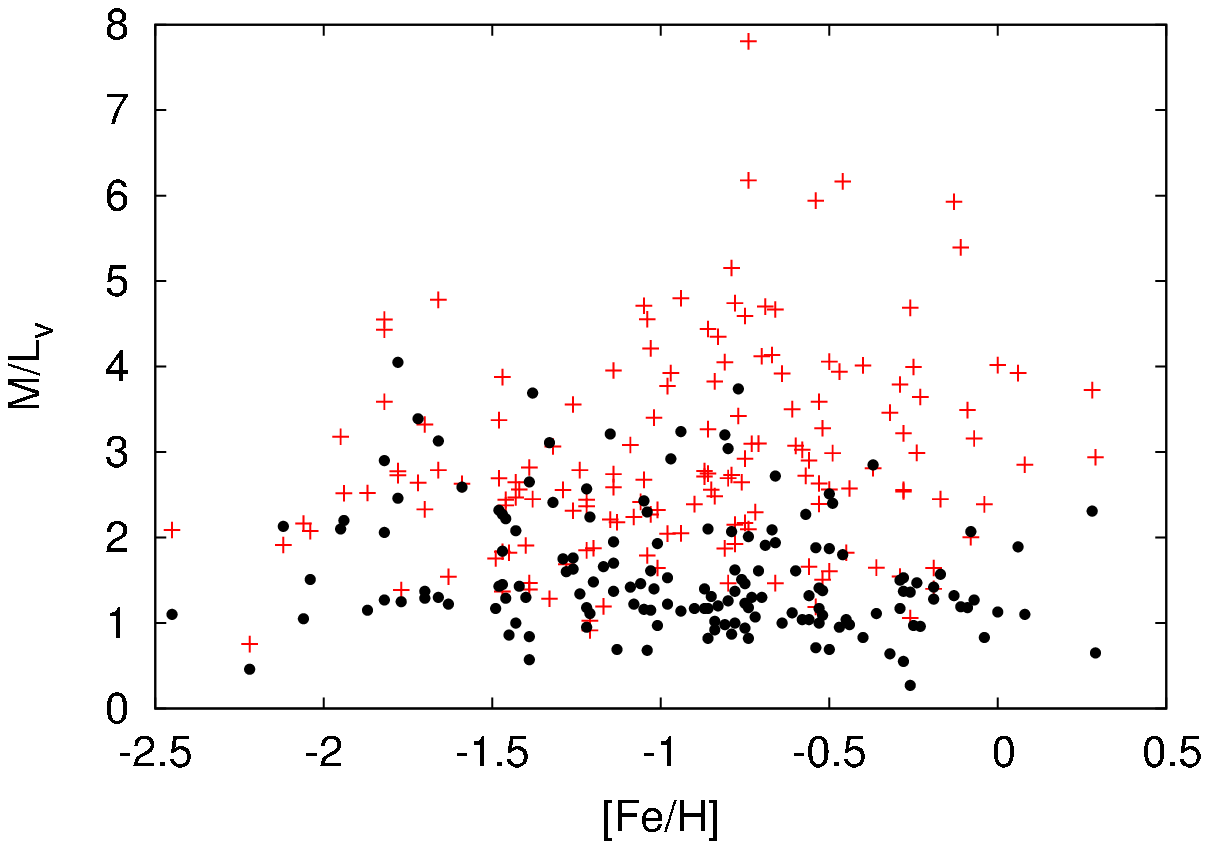}
\caption{The same as Fig.  3, but assuming a cluster-mass dependent retention fraction of remnants (i.e., BHs and NSs). See text for more details. } \label{mass-dependeny}
\end{figure*}

According to Fig. \ref{top-strong},  metal-rich GCs still typically show lower $M/L_V$ values than expected from SSP values. This discrepancy could be explained by the bias in the inferred mass from the integrated light properties (as a result of the assumption that mass traces light) that underestimates the true mass, especially at high metallicities as recently discussed  by \cite{Shanahan15}.

Another possibility could that the higher main sequence turn-off mass of metal-rich GCs (for clusters of the same age) leads to less mass in remnants (Sipple et al 2012). For instance, for an initial stellar mass of 50 $M_{\odot}$, the maximum black hole mass is about 28 $M_{\odot}$ for a metal-poor ($[Fe/H]\simeq -2$) progenitor and it is about 12 $M_{\odot}$ for a metal-rich ($[Fe/H]\simeq 0$) star (Belczynski et al. 2006, Sippel et al. 2012).  This suggests that the lower observed $M/L_V$ ratios of  metal-rich GCs in M31, may be a result of the smaller contribution of remnants in high-metallicity clusters, and also a spread in the retention fraction of NSs and BHs.

So far, we assumed that the retention fraction is the same for all clusters, while this fraction could vary in principle with the cluster escape velocity. In fact, for a given kick velocity dispersion of remnants the retention fraction of each remnant type depends on the local escape velocity, which is related to the cluster mass and radius as $V_{esc}= \sqrt{2GM/r_h}$. In order to estimate the dependency of the retention fraction on the cluster mass  the mass-radius relation from Larson (2004, $r_h\propto M^{0.1}$) or Marks \& Kroupa (2012, eq. 5 here) can be used. Regardless of the exact form of the mass-radius relation it can be easily shown that $V_{esc}\propto M^\alpha$, where $\alpha\simeq0.4$. Therefore, for a given kick velocity dispersion, the remnant retention fraction is set by the cluster mass, in a way that stellar remnants can be retained in massive clusters and therefore can have a significant impact on the cluster evolution affecting the M/L ratios. Since the exact form of the kick velocity distribution is still a matter of debate, in order to show the influence of the cluster-mass dependency of the retention fraction on the $M/L-[Fe/H]$ curve we assume the retention fraction to be changed linearly from zero (for the lightest massive cluster in our sample) to 0.7 (for the most massive cluster in our sample) as a function of the cluster initial mass. Fig. \ref{mass-dependeny} shows that the change is only marginal and does not affect the nature of the conclusions.

\begin{figure*}
\centering
\includegraphics[width=80mm]{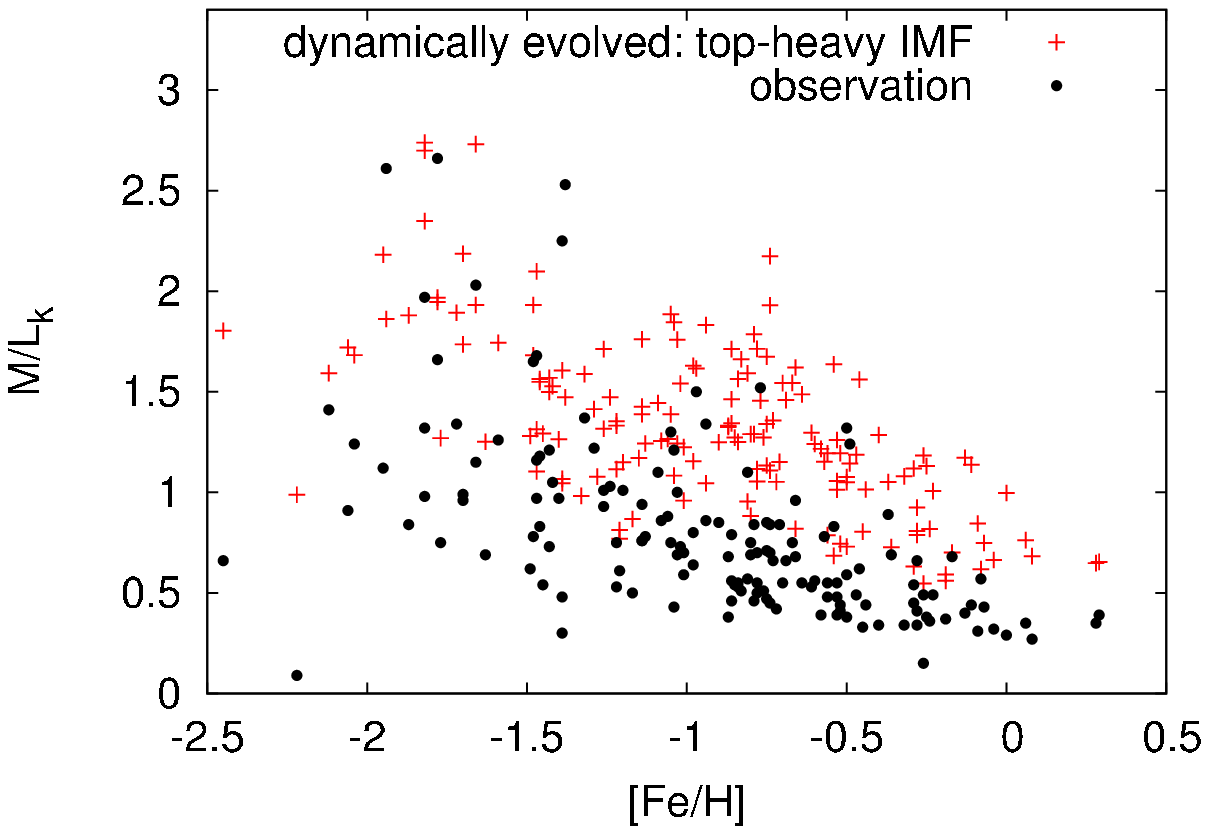}
\includegraphics[width=80mm]{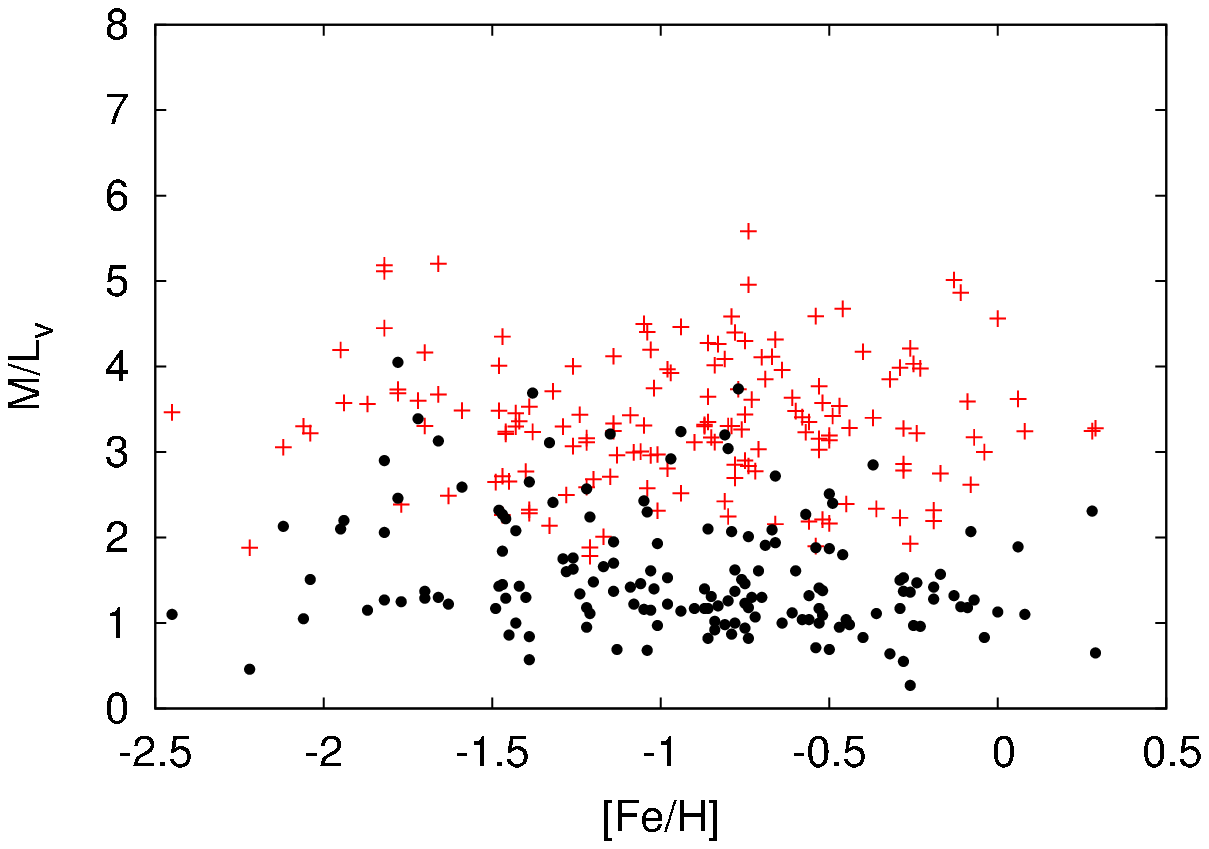}
\caption{The same as Fig.  3, but assuming the metallicity-dependent retention fraction of remnants (i.e., BHs and NSs; Eq. 7). See text for more details. } \label{top-metal-dep1}
\end{figure*}

\begin{figure*}
\centering
\includegraphics[width=80mm]{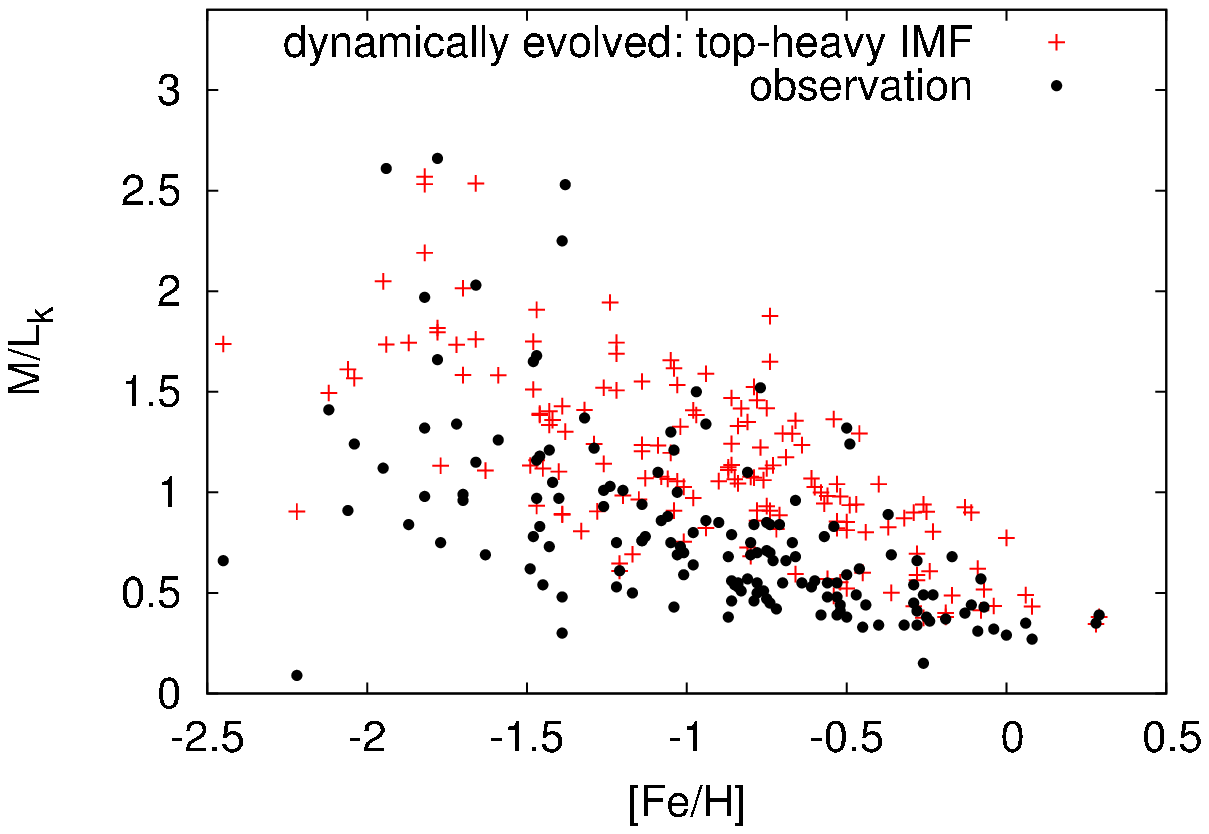}
\includegraphics[width=80mm]{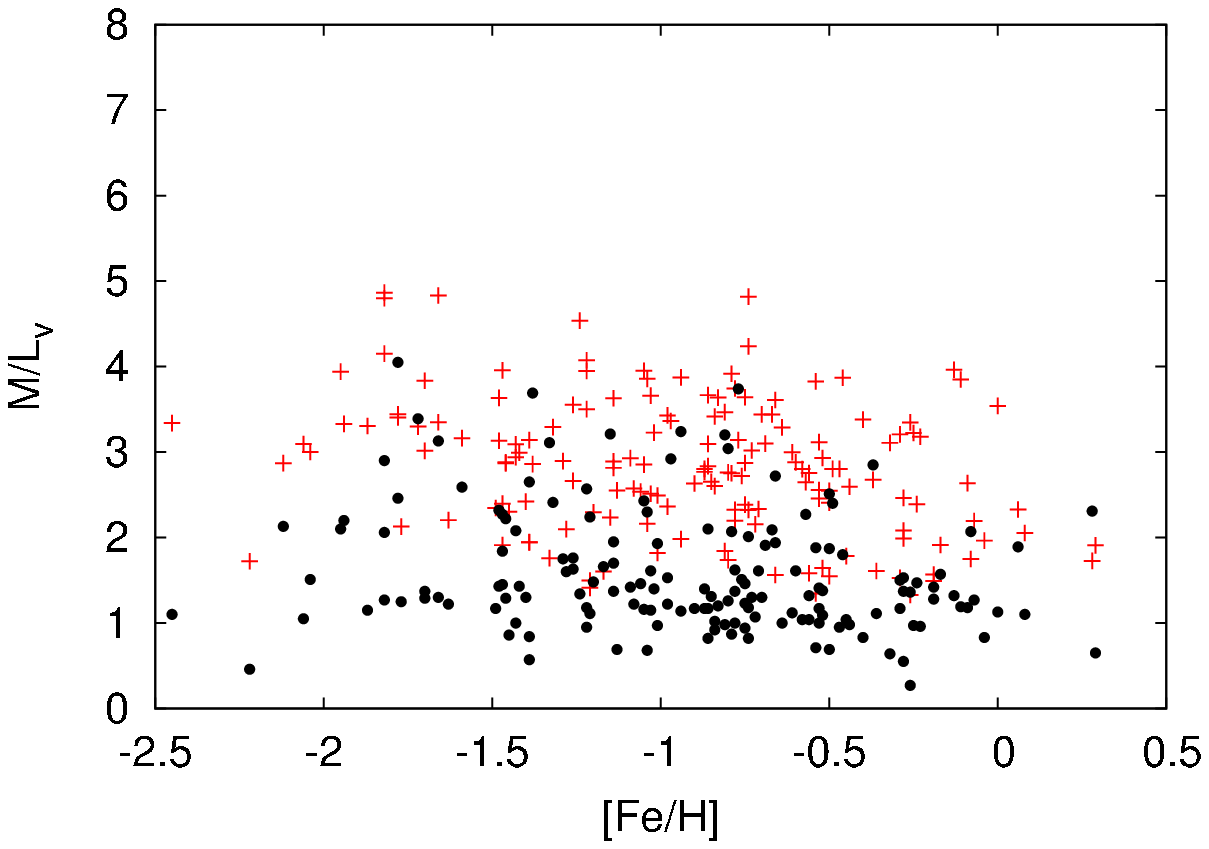}
\caption{  The same as Fig. 5, but assuming the same (Eq. 7) metallicity-dependent retention fraction for all remnants including WDs, NSs and BHs. } \label{top-metal-dep2}
\end{figure*}

As a final and perhaps most realistic model we adopt Eq. 4 for the variation of the IMF and additionally we assume the retention fraction of remnants ($r_f$) to be dependent on the metallicity,

\begin{equation}
r_f([Fe/H])=-0.16 [Fe/H]+0.08,
\end{equation}

\noindent such that the retention fraction of stellar remnants runs linearly from 0.4 at $[Fe/H]=-2$ to zero at $[Fe/H]=0.5$.  We furthermore assumed that 50\% of all white dwarfs (WDs) leave their star clusters due to dynamical evolution and stellar-astrophysical processes \citep{Fellhauer03}.  For each $[Fe/H]$ we calculate the corresponding retention fraction of BHs and NSs (Eq. 7) and  derive the inferred $M/L$ value of each GC in M31. As shown in Fig. \ref{top-metal-dep1}  such a model well reproduces the observed distribution of $M/L$ values of M31 GCs in both V- and K-bands.  Note that the agrement between our models and the data can be improved further if the WDs are also assumed to follow the same metallicity-dependent retention fraction as BHs and NSs (Fig. \ref{top-metal-dep2}).

\section{Conclusion}

Observations show a shallow decline of $M/L$ ratios in the V- and K-band with increasing metallicity for M31 GCs, while higher $M/L_V$ ratios are expected from  SSP models due to the evolution of both mass and luminosity. Also, a minimal dependency of $M/L_K$ on metallicity is expected from SSP models. In this paper we have presented a possible scenario to explain the discrepancy between the observed $M/L$ ratios of a large sample of GCs in M31 and those predicted by SSP models as found by Strader et al. (2011).  The main conclusions of these calculations can be summarized as follows:

\begin{enumerate}
 \item First, we added the effect of standard dynamical evolution to the result of SSP models calculated with a canonical IMF in order to investigate if such an effect can explain the growing difference between the $M/L$ ratios in the V- and K-bands with increasing metallicity for M31 GCs. The evolution of the stellar MF is computed by considering the fitting functions  derived by Baumgardt \& Makino (2003) based on comprehensive direct N-body experiments. We found that, although the dynamical evolution leads to a decrease of the $M/L$ ratio as a result of dynamical mass segregation and evaporation of low-mass stars from the star clusters in a tidal field of a host galaxy, this effect alone cannot describe the observed anticorrelation between  the $M/L$ ratios and  metellicity for the M31 GCs.

\item We next investigated the impact of a top-heavy IMF on the results. We used the recently derived top-heavy IMF as a function of the density and metallicity of embedded clusters, in which star formation leads to a more top heavy IMF in denser and metal-poorer pre-GC cloud-cores (Eq. 4), and showed that keeping 30\% of the remnants (i.e., BHs and NSs) within the clusters, the standard dynamical evolution can significantly reduce the discrepancy between the observed $M/L-[Fe/H]$ relation of M31 GCs and those predicted by such SSP models. We furthermore showed that assuming the retention fraction of remnants to be correlated with the cluster initial mass can remarkably improve the agreement of the calculated $M/L$ values in the V- and K-bands to the observation  by decreasing the predicted $M/L$ ratios. 

\item We showed that a stronger dependence of the top-heaviness on the metallicity (Eq. 5) can also improve the consistency between the theoretical $M/L$ values and the observed ones. This conclusion can be looked at from another perspective: If the observed $M/L-[Fe/H]$ anticorrelation of M31 GCs is a result of a top-heavy IMF, then this may be interpreted as a new constrain on the strength of the dependency of the top-heaviness of the IMF on the metallicity of the progenitor giant molecular clouds.

\item Finally and perhaps most realistically, we assumed the variation of the IMF as under point 2 above and in addition we assumed the retention fraction of stellar remnants depends on metallicity (Eq. 6). Such models reproduce the observed distribution of $M/L$ values of M31 GCs in the V- and K-bands best.
\end{enumerate}

It is worth pointing out that, although the lower than expected  $M/L$  values of the Milky-Way GCs can be explained by the depletion of low-mass stars as a result of dynamical evaporation (Kruijssen \& Miske 2009), here we show that this effect alone can not be the only reason for the observed $M/L-[Fe/H]$ anticorrelation of the M31 GCs, and our calculations suggest that a metallicity dependent top-heavy IMF might be necessary for the initial conditions of very massive star clusters. Such a dependency has been constrained by Mark et al. (2012).

We remind the reader that, with our approach we are making some simplifying assumptions. For instance, our computations of the  effect of dynamical evolution is based on some simplifying assumption on the average orbital parameters of the  M31 GCs, and  some rough estimations for the initial masses of the clusters from their present-day masses.
Investigating this problem for each individual GCs by direct N-body simulations or any other  faster $N$-body methods (e.g., the Monte Carlo method)  can provide an improved estimation of dynamical evolution on the $M/L$ ratios of GCs assuming a varying IMF and a possibly metallicity-dependent retention fraction of stellar remnants. It should be noted that in Baumgardt \& Makino (2003) the clusters moved through an external galaxy that followed a logarithmic potential. However, this potential is a good assumption for clusters moving in the outer part of the host galaxy ($Rg\geq10$ kpc), for the clusters orbiting in the inner part the effect of disk and bulge should be taken into account.  Moreover, in Baumgardt \& Makino (2003) all remnants that in principle have a significant impact on the evolution and dissolution rate of star clusters are assumed to be escaped.  Further simulations would be required in the next future to modify the results of Baumgardt \& Makino (2003) incorporating the dynamical effect of retained remnants and a more realistic model for galactic halo on the dynamical evolution.

As we mentioned above, since only the projected distance from the galactic center is available for each M31 GC, it is challenging to calculate their expected evolution due to the degeneracy in orbits of GCs and the resulting differences in tidal forces. Future kinematic data will enable us to more detailed estimates of the expected dynamical evolution of many of these objects by improving their structural parameters.

\begin{thebibliography}{99}
\bibitem[\protect\citeauthoryear{Adams \& Fatuzzo}{1996}]{Adams96} Adams F.~C., Fatuzzo M., 1996, ApJ, 464, 256

\bibitem[\protect\citeauthoryear{Baumgardt \& Makino}{2003}]{Baumgardt03} Baumgardt H., Makino J., 2003, MNRAS, 340, 227

\bibitem[\protect\citeauthoryear{Banerjee, Baumgardt \& Kroupa}{2010}]{Banerjee10} Banerjee S., Baumgardt H., \& Kroupa P., MNRAS, 402, 371
\bibitem[\protect\citeauthoryear{Banerjee \& Kroupa}{2011}]{Banerjee11} Banerjee S., Kroupa P., 2011, ApJ, 741, L12
\bibitem[\protect\citeauthoryear{Banerjee \& Kroupa}{2013}]{Banerjee13} Banerjee S., Kroupa P., 2013, ApJ, 764, 29
\bibitem[\protect\citeauthoryear{Banerjee \& Kroupa}{2014}]{Banerjee14} Banerjee S., Kroupa P., 2014, ApJ, 787, 158

\bibitem[\protect\citeauthoryear{Bastian, Covey \& Meyer}{2010}]{Bastian10} Bastian, N., Covey, K. R., \& Meyer, M. R. 2010, ARA\&A, 48, 339

\bibitem[\protect\citeauthoryear{Belczynski et al.}{2006}]{Belczynski06} Belczynski K., Sadowski A., Rasio F. A., Bulik T., 2006, ApJ, 650, 303

\bibitem[\protect\citeauthoryear{Breen \& Heggie}{2013}]{Breen13} Breen P. G., Heggie D. C., 2013, MNRAS, 432, 2779

\bibitem[\protect\citeauthoryear{Bruzual \& Charlot}{2003}]{Bruzual03} Bruzual, G., \& Charlot, S. 2003, MNRAS, 344, 1000

\bibitem[\protect\citeauthoryear{Caldwell et al.}{2009}]{Caldwell09} Caldwell, N., Harding, P., Morrison, H., Rose, J. A., Schiavon, R., \& Kriessler, J. 2009, AJ, 137, 94

\bibitem[\protect\citeauthoryear{Conroy \& Gunn}{2010}]{Conroy10} Conroy, C., \& Gunn, J. E. 2010, ApJ, 712, 833

\bibitem[\protect\citeauthoryear{Conroy et al.}{2009}]{Conroy10} Conroy, C., Gunn, J. E., \& White, M. 2009, ApJ, 699, 486

\bibitem[\protect\citeauthoryear{Dabringhausen, Kroupa, \& Baumgardt}{2009}]{Dab09} Dabringhausen J., Kroupa P., Baumgardt H., 2009, MNRAS, 394, 1529

\bibitem[\protect\citeauthoryear{Dabringhausen, Fellhauer, \& Kroupa}{2010}]{Dab10} Dabringhausen J., Fellhauer M., Kroupa P., 2010, MNRAS, 403, 1054

\bibitem[\protect\citeauthoryear{Dabringhausen et al.}{2012}]{Dab12} Dabringhausen J., Kroupa P., Pflamm-Altenburg J., Mieske S., 2012, ApJ, 747, 72

\bibitem[\protect\citeauthoryear{Dib, Kim, \& Shadmehri}{2007}]{Dib07a} Dib S., Kim J., Shadmehri M., 2007, MNRAS, 381, L40

\bibitem[\protect\citeauthoryear{Dib et al.}{2007}]{Dib07b} Dib S., Kim J., V{\'a}zquez-Semadeni E., Burkert A., Shadmehri M., 2007,
ApJ, 661, 262


\bibitem[\protect\citeauthoryear{Fellhauer et al.}{2003}]{Fellhauer03} Fellhauer M., Lin D.~N.~C., Bolte M., Aarseth S.~J., Williams K.~A., 2003, ApJ, 595, L53

\bibitem[\protect\citeauthoryear{Giersz \& Heggie}{1996}]{Giersz96} Giersz M., Heggie D.~C., 1996, MNRAS, 279, 1037


%
\bibitem[\protect\citeauthoryear{Haghi et al.}{2015b}]{Haghi15}
Haghi H., Zonoozi A.~H., Kroupa P., Banerjee S., Baumgardt H., 2015, MNRAS, 3872, 85.

%
%
%

\bibitem[\protect\citeauthoryear{Heggie \& Giersz }{2014}]{Heggie14} Heggie D. C., Giersz M., 2014, MNRAS, 439, 2459

\bibitem[\protect\citeauthoryear{Kruijssen \& Mieske}{2009}]{Kruijssen09} Kruijssen, J. M. D., \& Mieske, S. 2009, A\&A, 500, 785

\bibitem[\protect\citeauthoryear{Kroupa}{2001}]{Kroupa01} Kroupa P., 2001, MNRAS, 322, 231

\bibitem[\protect\citeauthoryear{Kroupa, Aarseth, \& Hurley}{2001}]{kroupa01b} Kroupa P., Aarseth S., Hurley J., 2001, MNRAS, 321, 699

\bibitem[\protect\citeauthoryear{Kroupa}{2002}]{Kroupa02} Kroupa P., 2002, Science, 295, 82

\bibitem[\protect\citeauthoryear{Kroupa et al.}{2013}]{Kroupa13} Kroupa P., Weidner C., Pflamm-Altenburg J., Thies I., Dabringhausen J., Marks M., Maschberger T., 2013, Planets, Stars and Stellar Systems, Volume 5: Stellar Systems and Galactic Structure. Springer-Verlag, Berlin.

\bibitem[\protect\citeauthoryear{Lada \& Lada}{2003}]{Lada03} Lada C.~J., Lada E.~A., 2003, ARA\&A, 41, 57


\bibitem[\protect\citeauthoryear{Mackey et al.}{2008}]{Mackey08} Mackey A.~D., Wilkinson M.~I., Davies M.~B., Gilmore G.~F., 2008, MNRAS, 386, 65

\bibitem[\protect\citeauthoryear{Maraston}{2005}]{Maraston05} Maraston, C. 2005, MNRAS, 362, 799

\bibitem[\protect\citeauthoryear{Marigo et al.}{2008}]{Marigo08} Marigo, P., Girardi, L., Bressan, A., Groenewegen, M. A. T., Silva, L., \& Granato, G. L. 2008, A\&A, 482, 883

\bibitem[\protect\citeauthoryear{Marks \& Kroupa}{2012}]{Marks12} Marks, M. \& Kroupa, P. 2012, A\&A, 543, A8

\bibitem[\protect\citeauthoryear{Marks et al.}{2012}]{Marks12b} Marks, M., Kroupa, P., Dabringhausen, J., \& Pawlowski, M. S. 2012, MNRAS, 422, 2246

\bibitem[\protect\citeauthoryear{Megeath et al.}{2016}]{Megeath16} Megeath, S.~T., Gutermuth, R., Muzerolle, J., et al., 2016, AJ, 151, 5

\bibitem[\protect\citeauthoryear{Morscher et al.}{2015}]{Morscher15} Morscher M., Pattabiraman B., Rodriguez C., Rasio F. A., Umbreit S., 2015, ApJ, 800, 9

\bibitem[\protect\citeauthoryear{Pflamm-Altenburg, Gonz{\'a}lez-L{\'o}pezlira, \& Kroupa}{2013}]{Pflamm13}Pflamm-Altenburg J., Gonz{\'a}lez-L{\'o}pezlira R.~A., Kroupa P., 2013, MNRAS, 435, 2604

\bibitem[\protect\citeauthoryear{Portegies Zwart, McMillan, \& Gieles}{2010}]{Portegies10} Portegies Zwart S.~F., McMillan S.~L.~W., Gieles M., 2010, ARA\&A, 48, 431

\bibitem[\protect\citeauthoryear{Shanahan \& Gieles}{2015}]{Shanahan15} Shanahan R.~L., Gieles M., 2015, MNRAS, 448, L94

\bibitem[\protect\citeauthoryear{Sippel \& Hurley}{2013}]{Sippel13} Sippel A. C., Hurley J. R., 2013, MNRAS, 430, L30

\bibitem[\protect\citeauthoryear{Sippel  et al.}{2012}]{Sippel12}  Sippel A. C., Hurley J. R., Madrid J. P., Harris W. E., 2012, MNRAS, 427, 167

\bibitem[\protect\citeauthoryear{Strader et al.}{2011}]{Strader11} Strader J., Caldwell N., Seth A. C., 2011, AJ, 142, 8

\bibitem[\protect\citeauthoryear{Strader et al.}{2009}]{Strader09} Strader J., Smith G. H., Larsen S., Brodie J. P., Huchra J. P., 2009, AJ, 138, 547


\bibitem[\protect\citeauthoryear{Vesperini \& Heggie}{1997}]{Vesperini97} Vesperini E. \& Heggie D. C., 1997, MNRAS, 289, 898

\bibitem[\protect\citeauthoryear{Weidner et al.}{2013}]{Weidner13} Weidner C., Kroupa P., Pflamm-Altenburg J., Vazdekis A., 2013, MNRAS, 436, 3309

%

\end{thebibliography}
\end{document}